\documentclass[letterpaper,11pt]{article}
\usepackage{graphicx}
\usepackage[margin=1in]{geometry}
\usepackage[authoryear]{natbib}
\usepackage[hidelinks,breaklinks]{hyperref}

\title{An optimizable scalar objective value cannot be objective\\
and should not be the sole objective}
\author{Isabel Kloumann and Mark Tygert}

\begin{document}

\maketitle

\tableofcontents

\section{Introduction}
\label{intro}

The morality of algorithms and their potential for bias and discrimination are important concerns. A popular approach to machine learning and artificial intelligence is via the numerical optimization of objective functions, and adapting such an approach to handle ethics could seem natural: with a hammer in hand, everything looks like a nail. The hammer of much artificial intelligence is the optimization of objective values, so some might like to treat morality solely through such objective functions. However, relying solely on the optimization of scalar objective values is fraught with unavoidable flaws when dealing with real people. Unfortunately, certain ideals in physics, economics, and operations research and optimization fall short in ethics. The purpose of this memo is to highlight obvious, well-known limitations that often get lost in the numbers and mathematical formalism. This memo focuses almost exclusively on descriptive social science, suggesting only a few normative prescriptions in passing. The goal here is to deal with reality as it is, rather than to contemplate what could be the ideal if we were to design society from scratch.

What could we do instead of relying solely on scalar objective values, instead of choosing and targeting a ``metric'' as the sole measure of success? We could emulate the legal and political system, being reasonably reactive and instituting prohibitions, rather than trying to be omniscient and instituting permissions and centralized designs. In the presence of uncertainty, disagreement, and corruption or adversarial activities, setting boundaries can be more robust than trying to optimize objectives within the feasible set of solutions (unless the number of boundary pushers is too great). The key is to react rapidly, reasonably, reliably, resolutely, and righteously.

The problems with relying solely on the optimization of scalar objective values range from the abstract issues of heterogeneity and complicated dynamical systems discussed in Sections~\ref{differences} and~\ref{dynamicals} to the complications of constrained, slow motion and multiple optimizers discussed in Sections~\ref{beyond} and~\ref{multiple} to the realities of uncertainty and psychology discussed in Sections~\ref{reality} and~\ref{scientism}. Sections~\ref{judicial} and~\ref{conclusion} very briefly sketch some possible ways forward. One could argue that any one of the objections discussed below should preclude the optimization of scalar objective values from being good on its own; all together, the objections should be conclusive. Hopefully this addresses an analogue of the droll complaint to Andrei N.\ Kolmogorov reported by~\cite{kendall}: ``You [Kolmogorov] have supplied one proof of your thesis, and in the mathematics that you study this would perhaps suffice, but we historians prefer to have at least ten proofs.'' The basic theme is that the optimization of scalar objective values is insufficient for real people, so (at least insofar as sophisticated algorithms must work and plan around people) the optimization of scalar-valued objective functions cannot possibly be sufficient for ensuring the algorithmic fairness we desire.\footnote{\label{degenerate}Needless to say, for a sufficiently simple problem, optimizing a scalar objective function may be sufficient for a good algorithm to solve that problem. For instance, an algorithm controlling an autonomous vacuum cleaner for floors could integrate well into the home environment despite having only a very simplistic model of morality. In contrast, more sophistication is probably necessary for some of the algorithms powering tech behemoths. Furthermore, this memo considers only sufficiently simple, ``optimizable'' objectives; the memo does not consider any objective that is defined only a posteriori, defined in terms of the results of its own optimization. After all, what use is an objective function which is defined (almost circularly) in terms of its own consequences after optimization? For example, anything can be formulated as maximizing the function defined to take the value 1 for what actually happens and to take the value 0 for every other possibility, but that function is not ``optimizable'' in any useful sense.}

\section{Diversity}
\label{differences}

Sections~\ref{perspectives} and~\ref{values} below review two oft-neglected philosophical truisms, which are central to the discussion; Section~\ref{consequences} sketches some of their consequences. Section~\ref{treatmentVSequity} then highlights a particularly important consequence. The technical literature refers to the observations of Sections~\ref{perspectives} and~\ref{values} as ``epistemological pluralism,'' as discussed by~\cite{turkle-papert}.

\subsection{Different perspectives}
\label{perspectives}
Different perspectives are useful, with potentially consistent, complementary conclusions, even when the premises used to derive those conclusions may be contradictory. The key is to amalgamate the end products of multiple analyses, without needing to reconcile the hypotheses and other means used to obtain those ends.

The classic, uncontroversial example of this from mathematics is the foundations of differential and integral Calculus. The mainstream foundation assumes that there is no positive number less than every finite positive number epsilon; this assumption leads to the delta-epsilon approach of Augustin-Louis Cauchy and Karl Weierstrass (which is related to the ``method of exhaustion'' of Archimedes). An alternative foundation assumes that there do exist positive numbers --- infinitesimals --- less than every finite positive number; this assumption leads to the ``non-standard analysis'' of Abraham Robinson. Both approaches rigorously yield Calculus, though formulating solutions to practical problems is sometimes easier with one of the approaches (which approach is best varies according to the application) \dots\ and the combined results of applying the two approaches offer complementary insights (which is especially helpful when discretizing a continuous formulation in order to perform calculations on a digital computer). The collection of~\cite{arkeryd-cutland-henson} elaborates.

Section~\ref{consequences} below touches upon some consequences for fairness.

\subsection{Different values and localized behaviors}
\label{values}
Different people are different, not only in their values, but also in their behavior conditioned on their values. The extent and sense in which someone is rational or irrational varies from person to person. At the very least, some people are smarter than others, and the range of smarts is actually quite large. Moreover, people generally optimize locally, not globally, leading to ``hysteresis'' (``hysteresis'' is a technical term for ``history-dependent behavior''). For instance, the United States of America has yet to fully adopt the metric system which constitutes the International System of Units; the United States' idiosyncratic customary, traditional system of weights and measures persists largely by maintaining continuity in the historical milieu as an established standard, and deviations from that standard can be painful, whether or not the standard is especially sensible in terms of global optimality. Similarly, people often optimize locally in time (short-term optimization) rather than globally (long-term optimization); taking advantage of delayed gratification is not everyone's forte. The following Section~\ref{consequences} discusses some consequences for fairness.

\subsection{Consequences of these differences}
\label{consequences}
The above-mentioned properties of differences imply that leveraging such diversity can be powerful and that considering such differences is necessary for success. Taking into account the fact that different people are different is essential, whereas homogenization of their axioms and beliefs would be beside the point (and pointless) --- inconsistency of their mindsets is irrelevant so long as their outcomes are mutually agreeable. For example, mathematicians and physicists have long co-existed productively, despite the latter's thinking being so sketchy (well, physicists may object to their thinking being characterized as sketchy, but mathematicians think they know better). Moreover, many computer scientists consider the main purpose of the fast Fourier transform to be the accelerated multiplication of large integers (whereas applications to spectral analysis and spectral methods, such as accelerated convolutional filtering and the numerical solution of differential equations, are probably foremost in most others' minds). A less facetious example is the synergy of abstract theoreticians and concrete experimentalists complementing each other. Really, as transformatively useful as the work of~\cite{sutton-barto} and many others has proven in many situations, unification of everyone into some scalar ``utility'' value, driven by the control theory or reinforcement learning of scalar rewards, turns out to be impossible, unnecessary, and undesirable (due to these properties of differences). Fairness is not just a number.

\subsection{Equity in treatment versus equity in outcome}
\label{treatmentVSequity}
There is no single right answer to what bias and discrimination may mean, or at least no single right answer with which everyone agrees. Some prefer equity in ``treatment''; others prefer equity in ``outcomes.'' ``Treatment'' refers to society's rules governing its members; for instance, society may prefer that laws not explicitly discriminate between a northerner and a southerner. ``Outcomes'' refers to what society's members end up achieving; for example, society may prefer that northerners make the same on average as southerners. Furthermore, the line between ``treatment'' and ``outcome'' is murky: not everyone is the same, with the same access to resources and opportunities, endowed with the same history and abilities. Such differences can be used to argue for either equity in treatment or equity in outcomes, as can various levels and varieties of corruption and of checks and balances. Indeed, those benefiting from inequality in outcomes often argue for equity in treatment, whereas those benefiting from inequality in treatment often argue for equity in outcomes. Unfortunately, the different notions of equity are generally incommensurate, as detailed, for example, by~\cite{pleiss-raghavan-wu-kleinberg-weinberger} and (without all the formal mathematical proofs) by Section B.3
of~\cite{belmont}. There needs to be some sort of balance.

\section{Dynamical systems}
\label{dynamicals}

Sections~\ref{complexs} and~\ref{fittest} present people and societies as dynamical systems, with Section~\ref{complexs} discussing how the real dynamics cannot be formulated as tractable optimizations, aside from the trivial tautology discussed in Section~\ref{fittest}. Sufficiently simple dynamical systems sometimes admit formulations as tractable optimizations; however, the differences discussed in Section~\ref{differences} above quash any hope for the existence or desirability of such a formulation.

\subsection{People and peoples are complex systems}
\label{complexs}
Really, humans are complicated dynamical systems playing the game of life, not optimizing any single scalar-valued objective over time.\footnote{\label{physics}Being part of reality and thus governed by the laws of physics, the closed system constituting the whole universe does always exist at the locally stationary points of the Lagrangian action, at least to the extent that the time evolution is deterministic, as discussed by~\cite{arnold} and~\cite{wald}. Yet humans do not constitute any kind of closed system. Moreover, the very concept of entropy and its maximization depends on a rather arbitrarily defined probabilistic, statistical, or fine-to-coarse multiscale model; notions such as Kolmogorov complexity can avoid some of the arbitrariness in the probabilistic model, but such notions are incomputable (and therefore incomprehensible as principles for falsifiable scientific prediction), as discussed in connection with the second law of thermodynamics (entropy maximization) by~\cite{cover-thomas}.} Society is an even more complex dynamical system. Here, ``dynamical system'' refers simply to how everyone and everything consists of many interacting parts acting rather reflexively and often stochastically to stimuli. For instance, anyone with exquisitely good taste can understand the modern-art and fashion communities only as dynamical systems, not optimal in any way, to the extent that they make any sense at all. Joking aside, formal definitions have been elaborated, for example, by~\cite{bertalanffy} and~\cite{boulding}. Given the complicated dynamics, designing a complete set of understandable laws or algorithms to be consistent, much less optimal according to an agreeable scalar objective value, must be utterly intractable. Different cases require different balances of people's rights.

\subsection{Survival of the fittest and their supporters}
\label{fittest}
Regarding aspirations toward optimality, the nonexistence of any reasonable scalar objective means that the only notion which makes sense for a very complicated dynamical system or game (in the game-theoretic sense) in the long run is ``survival of the fittest,'' where ``fittest'' means ``whatever survives.'' (Yes, that is circular and tautological, as caricatured by~\cite{dawkins}. See also Footnote~\ref{degenerate} in Section~\ref{intro} regarding objective functions defined retrospectively, that is, in hindsight.) The complex dynamics is rather arbitrary and subject to strong dependence on artifacts of history. Admittedly, stability may entail adherence to the basic ethical principle of no exploitation: individuals are responsible for supporting those who make their lifestyle possible. Yet such a principle is only a limited constraint, not fully specifying the behavior. For instance, while extraordinarily commandeered income or wealth inequality may lead to conflict, restricting to more equitable incomes and wealth is only a mild restriction. The non-uniqueness of optima discussed in Section~\ref{pareto} is related. Survival (with or without exploitation) fails to fully specify values being optimized, except after the fact.

\section{Optimization beyond the objective value}
\label{beyond}

Sections~\ref{hysteresis} and~\ref{constrainers} review how the objective value is often less important than other aspects of optimization, such as the dynamics discussed in Section~\ref{hysteresis} and the constraints discussed in Section~\ref{constrainers}. That dynamics and the constraints on those dynamics are important should be unsurprising, given the observation in Section~\ref{dynamicals} above that people and peoples are complex dynamical systems.

\subsection{Hysteresis due to local rather than global optimization}
\label{hysteresis}
Maximizing a scalar objective value is often visualized as riding up and down (but mostly up) the hills and valleys of a landscape, trying to get as high as possible (viewing altitude as the objective value being maximized). The latitude and longitude are coordinates which describe the ``state of the world'' or ``allocation of resources''; to fully describe the whole world requires more than just two numbers (latitude and longitude), of course, but visualizations usually restrict to a small number of dimensions \dots\,anyway, the analogy for a large number of dimensions should be clear. A classic problem in optimization is the tendency to try moving up always, getting stuck at the peaks of small hills rather than sometimes descending in order to reach and ascend larger mountains. Many of the methods for optimization discussed by~\cite{press-teukolsky-vetterling-flannery} target this problem of finding local rather than global optima.

People and societies do tend to get stuck at local optima. The landscape for optimization of those who are optimizing is highly non-convex, that is, there are many hills and valleys of varying heights and depths. Indeed, this non-convexity arises for many reasons, including the nonexistence of convex utility functions for all individuals (discussed below) and the presence of ``market failures'' that can block attainment of a fully efficient equilibrium, creating distortions and hurdles too high to be overcome. Market failures range from ``moral hazard'' and ``information asymmetry'' (for example, salespeople can profit from selling ``lemons'' --- defective products whose defects afflict only the purchasers) to the ``tragedy of the commons'' (overfishing is a perennial problem, for example) to ``externalities'' (pollution is a classic example) to ``outdated best practices'' (bloodletting took millennia to die) to inviolable ``collusion'' (try selling nuclear weapons for other than the government's asking price, for instance) to high ``barriers to entry'' (try starting a copycat competitor to the national postal system, for example). And, as discussed by~\cite{tinbergen}, this is only a small selection of the complications.

Randomness helps ameliorate the tendency to get stuck, yet ``change is hard,'' ``conformity'' (\cite{forsyth}), ``cognitive inertia'' (\cite{good}), ``institutional inertia'' (page 76 of~\cite{tinbergen}), ``myopic decision making'' (\cite{hinson-jameson-whitney}), and ``systemic or systematic risks'' (\cite{hull}) are all commonly discussed phenomena that must be overcome to attain substantial progress. As discussed by~\cite{hinson-jameson-whitney}, few are willing to sacrifice in the short term so as to attain great long-term gains, and fewer still are able to effect the massive broadly coordinated changes often required to move down the molehill surrounding a local maximum and then move up a mountain. Focusing solely on the objective value disregards these necessarily complex dynamics.

\subsection{The overwhelming influence of constraints}
\label{constrainers}
Optimizing locally seldom optimizes globally. As reviewed by~\cite{stiglitz-walsh}, classical economics often considers the world ``ceteris paribus'' --- ignoring how constraints such as existing capital development and distribution, technology, social structures, etc.\ come about. However, those constraints usually do more to affect the outcome than any optimization within the constraints. Who doubts that the prominence of Coca Cola has much to do with development of its brand? And what is more important to wireless communications than control of the airwaves or the technology to use them? Without the development of modern technology, we would still mostly be farmers, if not hunter-gatherers. Another example is how the state of medicine and regulations influence the practice of individual medical doctors more than any decisions they get to make individually. Yet setting the constraints well does require coordination of efforts across individuals, especially to achieve global rather than local optima. Signaling via prices to communicate the necessary information requires markets, yet markets seldom make themselves, and nearly never make themselves perfect or complete, as emphasized by~\cite{stiglitz-walsh}. The constraints on optimization or on more general dynamics tend to be more influential than any quantity being optimized. It is essential that the boundaries we set for ourselves be fair, even if that requires the boundaries to be exquisitely elaborate.

\section{Multiple decision-making agents}
\label{multiple}

Sections~\ref{pareto} and~\ref{james-stein} highlight the complexities of collecting together multiple decision-making agents, as in societies. Section~\ref{pareto} discusses how efficiency is nice but not the whole story. Section~\ref{james-stein} reviews a fundamental drawback of combining multiple criteria into a single scalar objective value. These sections elaborate on the dynamical properties discussed in Sections~\ref{dynamicals} and~\ref{beyond} above, focusing on the complexities of systems with multiple interacting optimizers. Sections~\ref{pareto} and~\ref{james-stein} are more technical than the rest of the memo and mainly target readers trained in economics, game theory, and statistics (or econometrics).

\subsection{Non-uniqueness of Pareto optima}
\label{pareto}
A ``Pareto optimum'' is an equilibrium state of the world for which there is no way without going against someone else's wont that anyone would change on his or her own volition. If everyone is trying to do better for himself or herself (whatever ``better'' is supposed to mean), then a Pareto optimum is when no one can do better without making someone else do worse. While Pareto optima can be preferable to other equilibria (and appropriate exchanges and trades may in principle move the world to such optima), they are not unique. In the technical terms used by economists when discussing ``Edgeworth boxes,'' choosing among the points along the Pareto frontier answers the question of ``equity'' or ``fairness,'' which is exactly the question at hand. Suffice to say that opting for Pareto optima is an incomplete solution; for more information, the discussion of~\cite{stiglitz-walsh} goes into far greater detail than possible here. And further complicating matters is the existence of one of the authors of this memo, as his purpose in life is to make everyone worse off.

\subsection{James-Stein shrinkage}
\label{james-stein}
Combining everyone and everything into a single intelligible scalar objective is inherently problematic. Such combinations immediately encounter mathematical problems even in the simplest of scenarios. For example, a classic result of~\cite{james-stein} (based on earlier work of Charles Stein) provides an estimator of the means for three or more independent unit-variance normal distributions from given independent and identically distributed samples, and the expected squared Euclidean distance of these estimates from the correct means is strictly less than the expected squared Euclidean distance of the canonical estimates from the correct means (the canonical estimates are the empirical means, each estimated independently of the means for the other independent normal distributions) --- uniformly no matter what the correct means happen to be. Thus, even though the normal distributions are independent, as are the samples from these distributions, the performance of an estimator of their means improves (no matter what the means happen to be) by making the estimate of each mean depend on the others \dots\,provided that ``performance'' is measured via a single scalar objective value, the expected value of the square of the Euclidean distance of the estimated means from their actual values. While the expected squared Euclidean distance of the estimated means from their actual values has got to be the most natural single number describing the performance of the estimates, optimizing this single scalar value leads to the bizarre consequence that combining totally independent data improves the estimates (when measuring the improvement via the single scalar objective value). The issue is obvious: combining independent errors into a single scalar in order to define optimality can reward combining independent statistics that have nothing to do with each other. One scalar is not enough.

\section{A real individual goes beyond optimizing a scalar}
\label{reality}

Sections~\ref{paradoxes} and~\ref{irrationalutilities} discuss actual human behavior, in the form of Allais' Paradox (in Section~\ref{allais}) and Ellsberg's Paradox (in Section~\ref{ellsberg}), as well as in terms of irrational behavior (with psychology discussed in Section~\ref{irrational}, the Utility Theorem of Von Neumann and Morgenstern discussed in Section~\ref{von-neumann-morgenstern}, and the inadequacy of reinforcement learning discussed in Section~\ref{reinforcement}). Thus, beyond the impediments to joint, global optimization discussed in Section~\ref{multiple} that are inherent to any complicated multi-agent system, in reality even the individual agents deviate from classical economists' ideals of optimizers. The presentation below follows economists' convention of distinguishing between quantifiable ``risk'' and unquantifiable ``uncertainty.''

\subsection{Two paradoxes}
\label{paradoxes}

\subsubsection{Maurice Allais' Paradox (uncertainty of risk)}
\label{allais}
For reasons of both tractability and the limitations of human reasoning, many decisions are made with somewhat fuzzy numbers, rather than strictly targeting a single well-defined precise scalar objective value. This is only natural, as illustrated by the following modernized ``paradox'' of~\cite{allais}:

Consider two separate choices, each between two gambles.

In one of the decisions (a), you must choose between getting \$1 billion for certain, or getting \$1 billion 89\% of the time, \$2 billion 10\% of the time, and nothing 1\% of the time.

In the other decision (b), you must choose between getting \$1 billion 11\% of the time (and nothing 89\% of the time), or getting \$2 billion 10\% of the time (and nothing 90\% of the time).

Almost everyone chooses getting \$1 billion for certain and getting \$2 billion 10\% of the time. Yet, if an individual has (finite) values $v_0$ for getting nothing, $v_1$ for getting \$1 billion, and $v_2$ for getting \$2 billion, then nearly everyone goes with (a) $v_1 > .89 v_1 + .10 v_2 + .01 v_0$ and (b) $.11 v_1 + .89 v_0 < .10 v_2 + .90 v_0$. But, inequality (a) is equivalent to $.11 v_1 > .10 v_2 + .01 v_0$ whereas inequality (b) is equivalent to $.11 v_1 < .10 v_2 + .01 v_0$, and hence the inequalities are contradictory. Perhaps most people round 89\% to 90\% and round 11\% to 10\% for decision (b); most people would not round the 1\% chance of nothing in decision (a) down to 0\% chance, of course. Do most people often have to make decisions in which the difference between an estimate of 89\% and an estimate of 90\% is statistically significant? If so, then they would have to be calculating very carefully! In the design of algorithms, overestimating people's adherence to certain ideals would not be really fair.

\subsubsection{Daniel Ellsberg's Paradox (uncertainty of uncertainty)}
\label{ellsberg}
Doing best according to a given model is hard enough for most; finding the best model is even harder, and in practice much is so completely unknown as to be unquantifiable. Real decision makers take this into account, as illustrated by the following modernized ``paradox'' of~\cite{ellsberg}:

Consider two separate choices, each between two gambles.

In one of the decisions ($\alpha$), you must choose between getting \$1 billion 95\% of the time or \$1 billion ($90+x$)\% of the time, where $x$ is some unspecified whole number between 0 and 10.

In the other decision ($\beta$) to be made, you must choose between getting \$1 billion 95\% of the time or \$1 billion ($100-x$)\% of the time (with the same $x$).

Flipping a fair coin determines which of the payoffs you get to take (that is, whether from ($\alpha$) or from ($\beta$)).

Many people strongly prefer to get the \$1 billion 95\% of the time for both decisions. Yet, if as proper ``Bayesian'' decision makers, we assigned a ``prior'' probability $p_x$ for each possible value of $x$, with the sum over $x = 0$, $1$, $2$, $\dots$, $10$ of $p_x$ being 1, then the expected value of the payoff when choosing both options whose success depends on $x$ would be the sum over $x = 0$, $1$, $2$, $\dots$, $10$ of $.5 (90+x)\% p_x v_1 + .5 (100-x)\% p_x v_1 = .95 v_1$, where $v_1$ is the value to the individual of winning \$1 billion. Yet $.5(95\%)v_1 + .5(95\%)v_1 = .95 v_1$ is also the expected value when choosing both options whose success is always 95\%, independent of $x$! Thus, no matter whatever fixed distribution of prior probabilities an individual could use, the proper Bayesian individual would be indifferent between the options involving and not involving the unknown $x$ (furthermore, learning the result of the flip of the fair coin would not update the Bayesian prior, as the coin has nothing to do with $x$). Many people apparently try to avoid unknown uncertainties. Why and how exactly they decide to avoid the unknown uncertainty remains an open question (do people mistrust their own reasoning in complicated scenarios?), but collapsing everything down to one precise number --- the ``expected risk'' --- is not what they do. Assuming that they do (as in much classical literature on economics and decision making) would be unfair.

\subsection{Irrationality and utilities}
\label{irrationalutilities}

\subsubsection{Irrational human psychology}
\label{irrational}
There are whole fields devoted to ``irrationality,'' including behavioral economics and many branches of cognitive and social psychology. Irrationality happens both at the level of the individual and in social phenomena such as the ``bandwagon effect,'' ``herd behavior,'' ``groupthink,'' ``cargo cults,'' and ``echo chambers.'' Recently these studies and terms have become part of popular science, too, including via works of~\citet{kahneman} and~\citet{thaler}. Related concepts include the ``bounded rationality'' and ``satisficing'' of~\cite{simon}. Whatever people do, they do not behave as perfectly rational, cold-hearted calculators, faultlessly greedily maximizing perceived utility. Around people, algorithms ought not assume too much rationality.

\subsubsection{(Non)existence of utilities in reality}
\label{von-neumann-morgenstern}
The Utility Theorem of~\cite{von_neumann-morgenstern} states that if the (counterfactually) elicited preferences of an individual satisfy four axioms (namely: completeness, transitivity, continuity, and independence of irrelevant alternatives), then there exists a scalar-valued ``utility'' function of the individual for which the individual's decision will maximize the expected value. For actual practice, few real individuals can make all the necessary numerical calculations in time for every decision in order to satisfy the axioms of the theorem (Allais' Paradox discussed in Section~\ref{allais} is a classic example of violating the independence of irrelevant alternatives). This is especially true when considering different time horizons and trying to discount utility appropriately. Moreover, unquantifiable ``uncertainty'' yields a different result than does quantifiable ``risk'' (as in Ellsberg's Paradox discussed in Section~\ref{ellsberg}), and uncertainty is greater when the time permitted for determining a preference and making a decision is limited.

Postulating the four axioms is suspect in many real-life scenarios. For instance, would someone really be ``irrational'' for preferring Princeton to Stanford, preferring Stanford to Yale, and preferring Yale to Princeton (perhaps while not being able to decide among all three simultaneously)? The preferences just mentioned violate the axiom of transitivity. Such preferences can easily arise via a ``Condorcet Paradox'': an individual may have many criteria for rating one university against another, and a majority of these criteria could prefer Princeton to Stanford, Stanford to Yale, and Yale to Princeton for each pairwise comparison (with a different majority in each comparison). Besides, in reality people do not consider counterfactually (hypothetically) every possibility, much less elicit their response to every possibility experimentally; any assumption of what would happen if they did is therefore not fully falsifiable. Furthermore, as also noted by~\cite{von_neumann-morgenstern}, people's preferences tend to change after every decision they make, complicating the measurement of ``revealed'' preferences or ``subjective'' probabilities (even if such did in fact exist).

Thus, the existence of the celebrated Utility Theorem should have little bearing on ethics or morality, as real people may not obey the theorem's assumptions.

\subsubsection{Just rewards?}
\label{reinforcement}
The lack of a good instantiation of what economists call ``utility'' (whether ``marginal'' or otherwise) precludes much real behavior from being dictated by ``value functions'' (or good incentives in the form of rewards), as in the ``reinforcement learning'' of~\cite{sutton-barto} and others \dots\,unless these are vector-valued rather than scalar-valued, not admitting a ``total'' ordering of their values. (A ``total'' ordering is a relation, conventionally denoted ``$\le$'', for which any two objects can be related: $a \le b$ or $b \le a$, for any objects $a$ and $b$, with $a = b$ if and only if $a \le b$ and $b \le a$; and for which $a \le b$ and $b \le c$ implies $a \le c$, for any objects $a$, $b$, and $c$. Collections of scalar numbers such as integers or real numbers have natural total orderings; collections of arbitrary sets of numbers rarely have natural total orderings.) A scalar value function can be only part of the whole picture.

\section{Scientism}
\label{scientism}

Optimizing objective values is also subject to more general hazards of quantitative analysis. Beware of fallible people ``adjusting'' or ``correcting'' the data and statistics. In practice, so-called ``instrumental variables'' and ``propensity scores'' are often used to fudge results according to the preconceptions, incompetence, or even self-interest of the investigator, as discussed (for both instrumental variables and propensity scores) by~\cite{luo-gardiner-bradley} and~\cite{winkelmayer-kurth}. As emphasized by~\cite{pocock-elbourne} and myriad others, such attempts to control for confounding factors are treacherous without proper randomized experiments, while randomized experimentation on actual people tends to be taboo.

Corruption and bias (whether conscious or not) plague essentially all real communities, including social, scientific, and statistical communities. The observations of~\citet{huff}, ``How to Lie with Statistics,'' tellingly highlight what is perhaps the foremost ``use'' of statistics, as does a statement attributed by~\citet{wainer-koretz} to Winston Churchill: ``\dots when I call for statistics about the rate of infant mortality, what I want is proof that fewer babies died when I was Prime Minister than when anyone else was Prime Minister. That is a `political statistic'.'' Spinning facts or figures is ubiquitous. Scientific consensus is hard to attain --- especially hard for delicate social matters subject to statistical fluctuations --- and is often suspect even when attained. Ethical systems need to be robust to bad objectives.

\section{Judicial and professional systems}
\label{judicial}

Perhaps we should move beyond optimizing a single scalar objective and instead focus on heuristics, intelligence, experience, history, and precedent? Starkly, the key to any well-functioning set of laws has always been the judiciary, which resolves the inevitable contradictions in the application of laws, via ``fair and reasonable'' ``interpretations,'' that is, making decisions beyond what the laws could specify, decisions to which all parties will accede. \cite{goodin} comprehensively reviews the history and outlook. It is impossible to specify everything in writing ahead of time, and intractable to make intelligible written prescriptions be even mildly consistent, much less optimal according to a single universal scalar objective.

Politics and law are all about checks and balancing. Balance is also key in the medical profession: a doctor cannot distill a patient's treatment into optimization of a single scalar ``health objective value'' analogous to a credit score. The doctor instead runs big batteries of tests; even a blood test reports many metrics. The doctor may treat the patient so as to move the test results into ``normal'' ranges, where ``normal'' often varies from patient to patient and from population to population and might even depend on the patient's medical history, lifestyle context, and special needs. Moreover, balance is essential --- keeping a patient's triglycerides in check may take a back seat to moving blood pressure to desirable levels, which may in turn be worth sacrificing in order to maintain reasonable levels of insulin in diabetics, for example, whereas sacrificing insulin levels to manage blood pressure and keep triglycerides in check might make more sense if the risk of stroke or cardiac arrest is high. Single-mindedly optimizing one scalar metric to the detriment of others is unacceptable, while amending a single scalar to account for all infinitely many possibilities is infeasible. The same is even more true for laws and ethics \dots\ and algorithms.

\section{Conclusion}
\label{conclusion}

To recapitulate, reconciling differences in axioms and beliefs is unnecessary so long as their consequences are mutually agreeable \dots\ which is good, as there are large differences between different people (both in their values and in their behaviors). In particular, diverse notions of fairness are all of legitimate concern; no single scalar-valued metric of morality is the be-all and end-all. People and peoples are complex dynamical systems exhibiting complicated dynamics that no feasible optimization of a scalar-valued objective function can even characterize, much less manipulate ethically. Often the dynamics and especially the constraints on those dynamics are more important than any idealized trend toward some ``eventual'' ``optimum'' (which is typically only locally ``optimal,'' at that). And, irrespective of these concerns, aggregating the objectives of independent individuals into a single scalar unavoidably introduces undesirable dependencies (as in James-Stein shrinkage from statistics). Moreover, in reality each individual fails to optimize any scalar objective value, if only since there is uncertainty in quantifiable ``risk'' and unquantifiable ``uncertainty,'' not to mention the lack of any total order for an individual's preferences and decision-making. Indeed, the whole field of behavioral economics focuses on such observed ``irrationality.'' A sophisticated algorithm interacting with people should not optimize just a scalar objective value, as if people were models of perfection in maximizing some scalar ``utility.'' On top of all that, optimizing a scalar objective value is subject to all the usual hazards of quantitative analysis, from frustrating complication to confusion to corruption to deception with statistics (whether intentional or not).

For all the reasons discussed above, the legal and political systems need to be full of potential conflicts that get resolved case by case by trusted judges --- they cannot be fully consistent, optimizing some single scalar objective. There is no good globally consistent scalar-optimizing system, at least none that people will deem acceptable. Perhaps the best we can manage is to be reactive and institute prohibitions, rather than proactively instituting permissions through consistent centralized planning and control. Predicting the evolution of the whole world and all its bad actors can be worth trying, but will probably end up being largely hopeless. The key could be to react extraordinarily fast and resolutely, solving problems immediately as they arise. In the presence of uncertainty or disagreement, setting boundaries may be more robust than trying to design and optimize objectives within the feasible set of solutions. Of course, we can try to make refinements and predictions, but should be aware of our limitations in understanding everything and foreseeing the future. Even so, artificial intelligence and data science could help identify problems that might suggest good rules to enforce. And we could presumably measure success in various ways using appropriate data analysis. Yet the world is a complicated dynamical system, a big game, that cannot optimize anything in particular, certainly not over time. Any scientific or technological apparatus must take into account both this and the hard reality of corruption and well-known psychological and sociological deficiencies. The real world is not what some believe to be ideal; good judicious morality requires more than simply trying to optimize a single scalar objective value.

\section*{Acknowledgements}
\vspace{-.5em}
We would like to thank Amos Elberg (who highlighted how the legal system is reactive, instituting prohibitions, rather than proactive, instituting permissions), Norberto Andrade, Marco Baroni, Tom Cunningham, Yevgeniy Grechka, Jon Guerin, Alex Peysakhovich, Michael Rabbat, and Larry Zitnick.

\bibliography{obval}
\bibliographystyle{plainnat}

\end{document}